\documentclass[trackchanges,twocolumn]{aastex701}

\begin{document}

\title{The Mid-infrared Emitting Jet in the Black Hole V404~Cygni in Quiescence}

\author[orcid=0009-0002-6989-1019, gname=Eric, sname=Borowski]{E. S. Borowski}
\affiliation{Department of Physics \& Astronomy, Louisiana State University, 202 Nicholson Hall, Baton Rouge, LA 70803, USA}
\email[show]{eborow1@lsu.edu}  

\author[orcid=0000-0003-3318-0223, gname=Robert, sname='Hynes']{R. I. Hynes} 
\affiliation{Department of Physics \& Astronomy, Louisiana State University, 202 Nicholson Hall, Baton Rouge, LA 70803, USA}
\email{rhynes@lsu.edu}

\author[orcid=0000-0002-4669-0209, gname=Qiana, sname=Hunt]{Q. Hunt}
\affiliation{Department of Physics and Astronomy, University of Lethbridge, Lethbridge, AB T1K 3M4, Canada}
\email{qiana.hunt@uleth.ca}

\author[orcid=0000-0003-3906-4354, gname=Alexandra, sname=Tetarenko]{A. J. Tetarenko}
\affiliation{Department of Physics and Astronomy, University of Lethbridge, Lethbridge, AB T1K 3M4, Canada}
\email{alexandra.tetarenko@uleth.ca}

\author[orcid=0000-0002-7092-0326, gname=Richard, sname=Plotkin]{R. M. Plotkin}
\affiliation{Department of Physics, University of Nevada, Reno, NV 89557, USA}
\affiliation{Nevada Center for Astrophysics, University of Nevada, Las Vegas, NV 89154, USA}
\email{rplotkin@unr.edu}

\author[gname=Tariq, sname=Shahbaz]{T. Shahbaz}
\affiliation{Instituto de Astrofísica de Canarias, E-38205 La Laguna, Tenerife, Spain}
\affiliation{Departamento de Astrofísica, Universidad de La Laguna, E-38206 La Laguna, Tenerife, Spain}
\email{tariqshahbaz1@gmail.com}

\author[orcid=0000-0003-3105-2615,gname=Poshak, sname=Gandhi]{P. Gandhi}
\affiliation{School of Physics \& Astronomy, University of Southampton, Southampton SO17 1BJ, UK}
\email{poshak.gandhi@soton.ac.uk}

\author[orcid=0000-0003-0976-4755, gname=Thomas, sname=Maccarone]{T. J. Maccarone}
\affiliation{Department of Physics and Astronomy, Texas Tech University, Lubbock, TX 79409, USA}
\email{thomas.maccarone@ttu.edu}

\author[orcid=0000-0003-3124-2814, gname=James, sname='Miller-Jones']{J. C. A. Miller-Jones}
\affiliation{International Centre for Radio Astronomy Research, Curtin University, GPO Box U1987, Perth, WA 6845, Australia}
\email{James.Miller-Jones@curtin.edu.au}

\author[orcid=0000-0003-3944-6109, gname=Craig, sname=Heinke]{C. O. Heinke}
\affiliation{Department of Physics, University of Alberta, CCIS 4–181, Edmonton, AB T6G 2E1, Canada}
\email{heinke@ualberta.ca}

\author[orcid=0000-0002-8808-520X, gname=Aarran, sname=Shaw]{A. W. Shaw}
\affiliation{Department of Physics and Astronomy, Butler University, 4600 Sunset Avenue, Indianapolis, IN 46208, USA}
\email{awshaw@butler.edu}

\author[orcid=0000-0002-7930-2276, gname=Thomas, sname=Russell]{T. D. Russell}
\affiliation{INAF – IASF Palermo, via Ugo La Malfa, 153, I-90146 Palermo, Italy}
\email{thomas.russell@inaf.it}

\author[orcid=0000-0001-6682-916X, gname=Gregory, sname=Sivakoff]{G. R. Sivakoff}
\affiliation{Department of Physics, University of Alberta, CCIS 4–181, Edmonton, AB T6G 2E1, Canada}
\email{sivakoff@ualberta.ca}

\author[gname=Phil, sname=Charles]{P. A. Charles}
\affiliation{School of Physics \& Astronomy, University of Southampton, Southampton SO17 1BJ, UK}
\affiliation{Astrophysics, Department of Physics, University of Oxford, Keble Road, Oxford OX1 3RH, UK}
\email{P.A.Charles@soton.ac.uk}

\author[gname=Efthymios, sname=Palaiologou]{E. V. Palaiologou}
\affiliation{University of Crete, Department of Physics, Voutes University Campus, GR-70013 Heraklion, Greece}
\affiliation{Institute of Astrophysics, Foundation for Research and Technology – Hellas, P.O. Box 1385, GR-71110 Heraklion, Greece}
\email{palaiolo@physics.uoc.gr}

\author[orcid=0000-0002-6446-3050, gname=Pablo, sname=Reig]{P. Reig}
\affiliation{Institute of Astrophysics, Foundation for Research and Technology – Hellas, P.O. Box 1385, GR-71110 Heraklion, Greece}
\affiliation{University of Crete, Department of Physics, Voutes University Campus, GR-70013 Heraklion, Greece}
\email{preig@ia.forth.gr}

\begin{abstract}

Observations of some quiescent black hole X-ray binaries have revealed an excess of mid-infrared (MIR) emission above that expected from their donor stars. In one system, V404~Cygni, this excess has been variously suggested to arise from the accretion disk, circumbinary material, or a compact relativistic jet. Here we present simultaneous James Webb Space Telescope (JWST), Atacama Large Millimeter/submillimeter Array (ALMA), and complementary multi-wavelength observations undertaken to resolve this uncertainty. We observed large-amplitude 21~$\mu$m variability on short timescales with JWST, particularly a dramatic flare which swiftly rose to $\approx2.4$~mJy, over 10 times the lowest observed MIR flux density. Similar variability was simultaneously observed from radio to X-ray wavelengths with other facilities throughout the campaign. This variability and the flat radio/mm/MIR spectral index ($\alpha = 0.04 \pm 0.01$) suggest that the MIR excess at and longward of 21~$\mu$m in V404~Cyg does not arise from the accretion disk or circumbinary material. Instead, the emission at 21~$\mu$m is dominated by synchrotron radiation from a jet which persists into quiescence. This result reinforces the ubiquity of the disk-jet connection in accreting black holes across a range of masses and accretion rates.

\end{abstract}

\section{Introduction} 

Binary systems in which a black hole intermittently accretes material from a low mass donor star are known as Black Hole X-ray Transients (BHXRTs). These transient systems exhibit extended periods of quiescence (years to decades) punctuated by short, bright outbursts (lasting from months to years) primarily arising from an increase in the mass accretion rate in an accretion disk \citep{bhbs}. During outbursts, BHXRT spectral states can be classified as ``hard'' and ``soft'' based on the shape of the X-ray spectrum, which depends on the ratio of higher-energy (hard) and lower-energy (soft) X-ray photons.

BHXRT outbursts typically begin with a transition from quiescence to a high-luminosity hard state and end with a decay from a low-luminosity hard state back to quiescence. The hard states are highly variable and show a power-law X-ray spectrum associated with a hot inner accretion flow \citep{bellonimotta}. Another characteristic of the hard states is a flat or mildly inverted radio spectrum (flux density $F_\nu \propto {\nu}^\alpha$, where $\nu$ is frequency and the spectral index $\alpha \ge 0$), interpreted as partially self-absorbed synchrotron radiation from compact relativistic jets \citep{bkjet,hjjet,fenderjet}. This flat spectrum continues up to a break frequency, where the jet becomes optically thin to the emitted synchrotron radiation and above which the spectral index steepens to $\alpha \approx -0.7$. Observations of some BHXRTs in quiescence also reveal radio emission with a flat or slightly inverted spectrum which suggests that synchrotron radiation from such an outflow is present during this state as well \citep{radiospec,gallo06,qjets}. 

While there are commonalities between the hard and quiescent states it remains an open question whether there exists a fundamental difference between these states or if quiescence is simply a very low-luminosity extension of the hard outburst states \citep{spectralevo}. As such, the study of BHXRTs in quiescence can reveal the differences between quiescent and hard states and probe inflow-outflow coupling around compact objects at low accretion rates \citep{xrbjets}. Additionally, relativistic jets launched from the supermassive black hole at the center of M87 have been spatially resolved, including during the Event Horizon Telescope campaign in which the observed X-ray luminosity of the core revealed a mass-scaled accretion rate comparable to quiescent BHXRTs \citep{ehtm87}. Given the mass-invariance observed in black hole accretion and outflows \citep{funplane,asabove}, this supports the persistence of jets into the quiescent state of stellar-mass black hole systems. In turn, studies of these stellar-mass systems illuminate our understanding of their supermassive counterparts.

Mid-infrared (MIR) emission is expected to be dominated by the donor stars in BHXRTs. However, observations of a few quiescent BHXRTs with the Spitzer Space Telescope revealed an excess of MIR emission above that expected from the donors \citep{cbdisk}. 
The origins of these excesses were unclear. The original study suggested they could arise from the accretion disk or circumbinary material heated by the donor star \citep{cbdisk}. It is thought that winds driven off the accretion disk during outbursts play a role in regulating mass accretion onto black holes, and such a wind was observed during the 2015 outburst of V404~Cyg \citep{2016Natur.534...75M}. In principle, these outflows could supply a reservoir of circumbinary material. The presence of circumbinary material would have profound implications for the angular momentum evolution of binaries \citep{chenpdot, xucbdisk, fastpdot}. Indeed, some X-ray binaries have been shown to have faster than expected orbital decays, including two of the systems in the Spitzer sample showing an MIR excess, A0620--00 and XTE~J1118+480 \citep{ghfastspiral,ghfastdecay}.

V404~Cyg is one of three quiescent LMXBs that were observed with Spitzer to exhibit a MIR excess \citep{cbdisk}. The system has a 6.47~day orbital period, among the longest in its class, and the donor is an evolved star \citep{1992Natur.355..614C, kharg}. Based on an orbital inclination of $67_{-1}^{+3}$~degrees and a mass ratio ($M_{donor}/M_{accretor}$) of $q = 0.06$, the primary component has been identified as a $9.0_{-0.6}^{+0.2}$~$M_\odot$ black hole \citep{kharg}. The distance to the system has been determined with radio parallax measurements to be $2.226\pm0.091$~kpc \citep{prabu23}. The size of a putative quiescent jet has been constrained to less than 1.4~au \citep{gallo07, jmj2009} at 22~GHz, with an upper limit of 1~au between the 8.4 and 4.8~GHz emission regions \citep{prabu23}.

\citet{cbdisk} reported that in V404~Cyg, unlike the other LMXBs in the Spitzer sample, the MIR excess could be accounted for by their model for a viscously heated accretion disk. This is because their estimates of the excess followed a $F_\nu \propto \nu^2$ trend expected for the Rayleigh-Jeans tail of a blackbody spectrum. \citet{gallo07} reanalyzed the Spitzer data and, based on their revised estimate of the 24~$\mu$m flux, suggested two alternative models which they found to be more viable. In the first model, a double-blackbody fit to the IR/optical spectrum, the properties of the secondary blackbody implied a larger size than the orbital separation, consistent with circumbinary material as was proposed for other objects in the first study. The second model, a single blackbody plus broken power-law fit to the radio/IR/optical spectrum representing the donor star and synchrotron emission from a jet, respectively, provided as good a fit as the double-blackbody model while also having the advantage of accounting for the observed radio spectrum. The authors suggested that a variability study could discriminate between these two possible origins for the MIR excess \citep{gallo07}, as the variability timescales of synchrotron radiation (seconds or less) are much shorter than those expected in the case of thermal emission from circumbinary material (minutes or more).

V404~Cyg has previously displayed substantial radio, optical, and X-ray variability in quiescence, including correlated variability during simultaneous multi-wavelength observations. Significant correlations of X-ray and optical variability have been observed on multiple occasions \citep{hynes04,hynes09}. A study of archival VLA and Very Long Baseline Array (VLBA) radio data spanning 24 years found a factor of a few variability on timescales of minutes to decades \citep{plotkin19}. This radio variability has not yet been definitively correlated with the X-ray activity \citep{hynes09, rana16}. To date, only one study, which tracked V404~Cyg as it transitioned back into quiescence after its 2015 outburst, found tentative evidence of correlation between X-ray and radio emission with a 15~minute delay \citep{2017ApJ...834..104P}.

We organized a multi-wavelength observational campaign to identify the source of the MIR excess in V404~Cyg. The primary aims of this study were to quantify the variability of the MIR emission and identify correlations with variability at longer wavelengths which are thought to arise from a jet. We chose the primary observatories specifically to accomplish these goals. NASA's newest flagship, the James Webb Space Telescope (JWST), was an essential choice due to its unparalleled sensitivity in the MIR. Though not designed to maximize temporal resolution, the timing capabilities of JWST far surpass those of Spitzer and allow for a measurement of the variability timescales of the MIR emission with sub-second precision \citep{2025AJ....169...21S}. Simultaneous Atacama Large Millimeter/submillimeter Array (ALMA) Band 3 (97.5~GHz) observations were crucial to search for correlated variability at longer wavelengths. Simultaneous multi-wavelength coverage from radio to X-ray allowed more comprehensive spectral and variability studies of the inflow-outflow coupling and maximized the scientific yield of this unique new dataset. We therefore extend previous such multi-wavelength campaigns \citep{hynes04, hynes09, rana16} by the inclusions of JWST and ALMA.

\section{Observations and data reduction}

\subsection{James Webb Space Telescope (JWST)}

We performed a time-series observation (TSO) of V404~Cyg with the Mid-Infrared Instrument (MIRI) on the JWST with the F2100W filter (pivot wavelength $\lambda = 20.795\ \mu$m, bandwidth $\Delta\lambda=4.58\ \mu$m) utilizing the \texttt{FASTR1} readout mode in two exposures of about two hours each. As the target is not an extended source and we wished to maximize temporal resolution, we chose to observe in subarray mode. Each exposure included 1185 integrations (duration 5.99~seconds) comprised of 20 individual groups (duration $\sim 0.3$~seconds). The first exposure began on 2023-10-14 UTC 19:59:14.32 and ended at UTC 22:03:27.74. The second exposure began on 2023-10-14 UTC 22:14:33.87 and ended at 2023-10-15 UTC 00:18:47.35. Thus, the full TSO spanned 4.14~hours.

A failure of the standard JWST calibration pipeline made the data products retrieved from the Mikulski Archive for Space Telescopes (MAST) portal unsuitable for analysis. Instead, we retrieved the raw data and processed it locally. We found that the first stage of the pipeline (detector-level calibrations) was flagging good pixels as outliers. These issues were resolved with the pipeline adjustments described in \citet{2025MNRAS.537.1385G}. The second stage of the pipeline was executed with no problems, and we performed our analyses on the resulting calibrated products.

Standard aperture photometry was performed on the calibrated images using the python package \texttt{photutils}. We chose to use a 4.07~pixel aperture corresponding to 50\% encircled energy to maximize the signal-to-noise ratio ($S/N$) and applied an aperture correction according to the \texttt{jwst\_miri\_apcorr\_0014.fits} reference file. Any integrations which were affected by cosmic rays in the source or background regions were discarded, resulting in the 5.99~second time resolution light curve presented in Figure \ref{fig:mwlc}(d). Additionally, we have developed a scheme to achieve higher temporal resolution light curves of JWST/MIRI data by reducing the images at the group level rather than the default integration level. This group-level reduction was also performed and from these data a light curve with $\sim 0.3$ second time resolution was generated. The $S/N$ at this time resolution was low, so these were binned to 3 groups, yielding a $\sim 0.9$ second time resolution light curve, shown in part in Figure \ref{fig:mwlc}(a). In addition to revealing variability on timescales not captured in the default integration-level reduction, this procedure also revealed a higher peak flux density during the intense mid-infrared (MIR) flare at the start of the JWST observation than was evident at the integration-level. The mean observed 21~$\mu$m flux density during the TSO was $\approx0.42$~mJy.

Following the TSO, MIRI imaging was performed with five filters utilizing the \texttt{FASTR1} readout mode and a 4-point dither to measure the shape of the spectral energy distribution (SED) in the MIR (see Table 1). The stage three calibrated products retrieved from the MAST portal were utilized for all but the F2550W measurement. Those data suffered from calibration pipeline errors, which were resolved in a similar manner to the time-series data. As a check, we performed the same calibration and analysis routine on the raw images in the other four filters, which produced results consistent with the stage three data products retrieved from the MAST portal. 

\begin{deluxetable}{cccccc}
\tablewidth{0pt}
\tablecaption{JWST MIRI Post-TSO Imaging}
\tablehead{
\colhead{Filter} & \colhead{$\lambda_{\rm pivot}$ ($\mu$m)}& \colhead{Bandwidth ($\mu$m)} & Start time (UTC) & Elapsed time (minutes)\tablenotemark{\scriptsize a} & \colhead{Flux Density (mJy)\tablenotemark{\scriptsize b}} \\
}
\startdata
F1280W & 12.810 & 2.47 & 2023-10-15 00:33:15.659 & 4.34 & $0.696\pm0.010$\\
F1500W & 15.064 & 2.92 & 2023-10-15 00:40:35.082 & 4.48 & $0.584\pm0.010$\\
F1800W & 17.984 & 2.95 & 2023-10-15 00:47:51.155 & 4.46 & $0.402\pm0.010$\\
F2100W & 20.795 & 4.58 & 2023-10-15 00:55:05.476 & 4.59 & $0.377\pm0.010$\\
F2550W & 25.365 & 3.67 & 2023-10-15 01:03:07.724 & 10.30 & $0.362\pm0.010$\\
\enddata
\tablenotetext{\scriptsize a}{Time between start and end of exposure, including dither time.}
\tablenotetext{\scriptsize b}{Dereddened, adopting $A_V = 4.0$ \citep{hynes09,cas93,2021ApJ...916...33G} and $R_V = 3.1$}
\end{deluxetable}

\subsection{Atacama Large Millimeter/submillimeter Array (ALMA)}

V404~Cyg was observed with 40 antennas of the ALMA 12 meter array (Band 3 with a central frequency of 97.5~GHz) over two executions: the first from 2023-10-14 UTC 21:31 to 21:51, and the second from 2023-10-14 UTC 23:07 to 2023-10-15 UTC 00:24 (project code 2023.1.01719.S). The antennas were configured such that the minimum baseline was 89.6~m and the maximum baseline was 8282.7~m, achieving an RMS of 0.01~mJy over 6.625~GHz and a mean beam size of 0.158~arcsec. The data were reduced using the standard pipeline within the Common Astronomy Software Application package (\texttt{CASA}) v6.6.1 \citep{2007ASPC..376..127M}. Within this procedure, J1924-2914 was used for the bandpass and flux calibrator for the first execution, J2232+1143 was used as the bandpass and flux calibrator for the second execution, and J2025+3343 was used as the phase calibrator for both executions. The calibrated data were then visually inspected to ensure all atmospheric lines and artifacts from instrumental effects were properly removed. To compute flux density time-domain variability, we pass the calibrated interferometric data through a custom python script \footnote[1]{\href{https://github.com/Astroua/AstroCompute_Scripts}{https://github.com/Astroua/AstroCompute\_Scripts}} that generates high time resolution light curves by performing a multi-frequency synthesis imaging with the \texttt{CASA} \texttt{tclean} task for each time bin and fitting the source flux density in the resulting image plane with \texttt{imfit}. We defined a bin size of 60~seconds and used a Briggs weighting scheme which produced the cleanest results. 

\subsection{Karl G. Jansky Very Large Array (VLA)}

VLA observations were awarded through Directors Discretionary Time (Project ID 23B-315). We observed from 2023-10-14 UT 19:44:05 to 2023-10-15 UT 00:39:16, obtaining 4$^{\rm h}$ 8$^{\rm m}$ (4.13~hours) on V404~Cyg. Observations were taken in C-band centered at 6 GHz, with 4 GHz of bandwidth. The VLA was in the process of moving configurations from its most extended A configuration to its most compact D configuration, resulting in a hybrid array where the northern and eastern arms still contained some antennas as far as $\approx20$~km from the array center (seven total across both arms; maximum baseline 34.5~km). All other antennas were in their D configuration positions ($<1$~km; minimum baseline 40~m). We used 3C~286 as the flux density and bandpass calibrator, and J2025+3343 as the secondary calibrator, which we observed every 8 minutes to solve for time-dependent complex gain solutions.

Data were calibrated using the VLA pipeline v6.4.1, and additional analysis used \texttt{CASA} v6.5.0 \citep{casateam}. We performed small amounts of additional flagging before proceeding to imaging with the \texttt{CASA} task \texttt{tclean}, using 2 Taylor terms to account for the wide fractional bandwidth. We also experimented with different {\tt robust} values using a Briggs weighting scheme to balance sensitivity with minimizing sidelobes from other sources in the field. Initial attempts resulted in poor image quality; our secondary calibrator was only 16.6' from our target, which, combined with the elongated synthesized beam from the hybrid array configuration casted sidelobes near V404~Cyg. 

We thus opted to exclude all seven `A configuration' antennas, leaving an array of 20 antennas in close to D configuration and a more circular synthesized beam. We then peeled J2025+3343 from the measurement set, following the steps outlined by \citet{deller15}, and described below.

We first changed the phase center of the measurement set to the position of J2025+3343, and we performed two rounds of self-calibration to determine directional-dependent gain solutions. In the first round, we applied phase-only gain solutions on 10~min intervals, and in the second-round we applied phase and amplitude gain solutions over 20~min intervals (in both instances, we integrated over all spectral windows when calculating gain solutions to improve $S/N$). We then created a final image with \texttt{tclean} to build a sky model. After masking out all other field sources in the model, we subtracted the remaining model (including only J2025+3343) from the measurement set using the task \texttt{uvsub}. We next inverted the complex gain solutions from our self-calibration and removed the calibrations specific to J2025+3343 from the measurement set, which may not be appropriate for the direction of V404~Cyg. Finally, we changed the phase center back to the position of V404~Cyg, allowing us to create images without the bright nearby source J2025+3343 and improving the image quality near V404~Cyg.

We then created a light curve of V404~Cyg by creating images every 5 minutes, at a reference frequency of 6.0~GHz and using \texttt{robust=0}. The peak flux density of V404~Cyg was measured using \texttt{imfit} in each image, and the root mean square (rms) noise was measured in a nearby source-free region of sky. We checked that V404~Cyg was a point source in all images, so we therefore forced a point source when performing the source fitting in \texttt{imfit} to measure the peak flux density. The rms improved from $\approx 0.2$~mJy~bm$^{-1}$ in images toward the beginning of our observation (when V404~Cyg was rising and closer to the horizon) and improved to $\approx 0.06$~mJy~bm$^{-1}$ toward the end as V404~Cyg approached zenith. 

\subsection{Gran Telescopio Canarias (GTC) HiPERCAM}

We obtained high-speed optical imaging of V404~Cyg on 2023-10-14 using HiPERCAM on the 10.4~m Gran Telescopio Canarias in La Palma, Spain under the Director Discretionary Time program GTC07-23BDDT. HiPERCAM uses dichroic beamsplitters to simultaneously image the custom-made Super-SDSS $u_{\rm s}$, $g_{\rm s}$, $r_{\rm s}$, $i_{\rm s}$, and $z_{\rm s}$ filters \citep{2021MNRAS.507..350D}. The CCDs were binned by a factor of 2 and windowed to four windows of $248\times248$ pixels each.  The instrument was oriented so that one window was centered on V404~Cyg and the other windows on a local standard star and comparison stars. We took 276638 images with an exposure time of 0.24068~s which resulted in a cadence of 0.24849~s. The conditions were very good and stable with a median seeing of 0.7~arcsec.

The data were reduced using the HiPERCAM pipeline \footnote[2]{https://github.com/HiPERCAM/hipercam}. A bias image was subtracted from each frame and flat field corrections were applied.  We extracted the count rates for each star using aperture photometry with a 2.3~arcsec circular aperture tracking the centroid of the source. Since there is a line-of-sight contaminating star 1.5~arcsec north of V404~Cyg, we determined the combined counts using an aperture which encompassed both stars.  The count ratio of V404~Cyg with respect to the local standard (2.05' northwest of V404~Cyg) was then determined by subtracting the count ratio of the contaminating star with respect to the local standard (determined from images taken under good seeing conditions $< 0.5$~arcsec) from the combined count ratio of V404~Cyg (i.e. V404~Cyg + line-of-sight star) with respect to the local standard.

For our DDT observations no standard star observations were performed so we instead used V404~Cyg and HiPERCAM standard star \citep{2022MNRAS.513.3050B} observations taken on 2024-09-30 to determine the instrumental zero-point, which was then used to calibrate the stars in the V404~Cyg field-of-view. Given that the local standard and comparisons stars are in the Pan-STARRS survey DR1 catalog \citep{2020ApJS..251....6M}, we transformed these to SDSS magnitudes \citep{2016ApJ...822...66F} and then compared them with the magnitudes determined using our own calibration. We found that they agreed at the $\sim 20\%$ level. Finally, we converted our SDSS magnitudes to flux densities. The mean observed magnitudes of V404~Cyg were $g_{\rm s}=20.25$, $r_{\rm s}=18.05$, $i_{\rm s}=17.03$, and $z_{\rm s}=16.10$. 

\subsection{Skinakas Observatory}

Optical monitoring of V404~Cyg was carried out at Skinakas Observatory in Crete, Greece with a 1.29~m (f/7.6) Ritchey-Chr\'{e}tien telescope from 2023-10-14 UTC 16:56:23 to 22:10:40 using a Cousins $R$ ($R_{\rm C}$) filter and a $2048\times2048$ back-illuminated, deep-depletion CCD camera (Andor iKon-L) with pixel size 13.5~$\mu$m. The system provides a $9.6'\times9.6'$ field-of-view and image scale 0.283~arcsec per pixel. Flat-field images were acquired during the evening twilight, and bias frames (30 in total) were taken before the start of the source monitoring and right after the end. The exposure time was set at 60~seconds, resulting in a $S/N$ of about 145--160. 290 images were collected at a mean cadence of 65.25~seconds. The total duration of monitoring was 5$^{\rm h}$ 14$^{\rm m}$ 17$^{\rm s}$ (5.238~hours). Sky conditions were photometric throughout the complete run and seeing was good; the FWHM (for a 60~second exposure) ranged from $\sim 0.8-1.2$~arcseconds, with most of the images having a FWHM in the sub-arcsecond regime. The target was at an airmass 1.002 at the start of the monitoring which ended when the airmass reached a value $\sim 2$. Raw images were trimmed (to exclude prescan and overscan columns), bias subtracted, and flat-field corrected using standard \texttt{IRAF} \citep{1986SPIE..627..733T} routines. Standard aperture photometry was performed using a source aperture large enough to encompass both V404~Cyg and the contaminating star. Using \texttt{daophot} \citep{1987PASP...99..191S} as implemented in \texttt{IRAF} we extracted instrumental magnitudes for V404~Cyg, the line-of-sight contaminating star 1.5~arcsec to the north, and four bright comparison stars nearby. The count ratio of V404~Cyg with respect to the local standard was determined by subtracting the count ratio of the contaminating star with respect to the local standard (2.05' northwest) from the combined count ratio of V404~Cyg + contaminating star with respect to the local standard. As the HiPERCAM observations were calibrated independently with standard star observations, we elected to calibrate the Skinakas data by applying an offset so that the mean flux density as measured by Skinakas matched the mean flux density as measured by HiPERCAM during the time that the observations overlapped. The mean observed magnitude of V404~Cyg was $R_{\rm C} = 17.87$.

\subsection{Chandra X-ray Observatory}

Chandra observations were awarded through Directors' Discretionary Time (Proposal ID 24408930). V404~Cyg was observed by Chandra from 2023-10-14 UTC 20:11:51 to 2023-10-15 UTC 00:57:51 for 15.0~ks of good time using the ACIS-S camera. We used the 1/8 subarray mode to reduce frame time to 0.4~s and ensure negligible pile-up. The target was positioned at the recommended aim-point on the S3 chip. The data were analyzed with \texttt{CIAO} \citep{2006SPIE.6270E..1VF} v4.16. We extracted source events with energies 0.3--7.0~keV from a 3.2~arcsec radius aperture and the background from a 23~arcsec circle near the target. We recorded a total of 0.04~counts~s\textsuperscript{-1} and estimate a background rate in the aperture of $5 \times 10^{-5}$~counts~s\textsuperscript{-1}. We fit the spectrum in \texttt{XSPEC12} \citep{1996ASPC..101...17A} with an absorbed power-law model (\texttt{wabs*powerlaw}) with the absorption column fixed to $N_H = 0.88 \times 10^{22}$~cm\textsuperscript{-2} \citep[adopted from][]{2007ApJ...667..427B}). We obtained a photon index $\Gamma = 2.14 \pm 0.11$ and an average unabsorbed flux of $1.5 \times 10^{-12}$~erg~cm\textsuperscript{-2}~s\textsuperscript{-1} corresponding to an average luminosity of $L_X = 8.9\times 10^{32}$~erg~s\textsuperscript{-1} at 2.226~kpc.

\subsection{XMM-Newton}

V404~Cyg was observed by XMM-Newton as a Target of Opportunity program (Revolution 4367 ObsID 0932390301) from 2023-10-14 UTC 17:13:21 to 20:54:59 for 13.7~ks of good time using all three EPIC cameras (EMOS1, EMO2, and PN). We used a medium filter with imaging full window mode. We reprocessed and analyzed the data using \texttt{SAS} \citep{2014ascl.soft04004S} version 21.0.0. We filtered events with \texttt{PATTERN} $\le 12$ cleaning for EMOS and \texttt{PATTERN} $\le 4$ for PN, together with an energy selection of 0.3--7.0~keV to maximize compatibility between instruments and with Chandra. For the EMOS cameras we used a 20~arcsec source extraction region with a 100~arcsec annular background region around the target. V404~Cyg was located near the edge of the chip in PN, so we used a 15~arcsec extraction radius with a separate circular background region of radius 68~arcsec. We recorded a total of 0.04, 0.03, and 0.10~counts~s\textsuperscript{-1} in EMOS1, EMOS2, and PN respectively, and estimate background rates in the aperture of 0.009, 0.006, and 0.03 respectively. We jointly fit the three spectra using the same model as for Chandra. We obtained a photon index $\Gamma = 2.23 \pm 0.08$ and an average unabsorbed flux of $9.5 \times 10^{-13}$~erg~cm\textsuperscript{-2}~s\textsuperscript{-1} corresponding to an average luminosity of $L_X = 5.7\times 10^{32}$~erg~s\textsuperscript{-1} at 2.226~kpc.

\section{Results}

Between XMM-Newton and Chandra X-ray Observatory, there was continuous coverage in X-rays for the duration of our observations. During this time, the average X-ray luminosity of V404~Cyg was $L_{0.3-7.0\ \rm keV} \approx 6.6 \times 10^{32}$~erg~s\textsuperscript{-1} which corresponds to $\sim 5.3 \times 10^{-7}\ L_{\rm Edd}$ for a 9.0~$M_{\odot}$ black hole. These measurements are within the range of X-ray luminosities associated with quiescent BHXRTs \citep{bhbs,spectralevo} and are consistent with V404~Cyg being among the most X-ray luminous quiescent Galactic BHXRTs \citep{2007ApJ...667..427B, 2014MNRAS.439.2771B}. The average X-ray luminosity observed with XMM-Newton before the start of the JWST observations was $L_{0.3-7.0\ \rm keV} \approx 2.6 \times 10^{32}$~erg~s\textsuperscript{-1}, corresponding to $\sim 2.1 \times 10^{-7}\ L_{Edd}$. This is among the lowest X-ray luminosities observed from V404~Cyg. This is a similar Eddington ratio as that observed during the Event Horizon Telescope campaign targeting the accreting supermassive black hole at the center of M87. In that campaign, the core X-ray luminosity (accretion flow plus inner jet) as measured by Chandra X-ray Observatory, which provides an upper limit on the accretion luminosity, was $L_{0.3-7.0\ \rm keV} \approx 9 \times 10^{40}$~erg~s\textsuperscript{-1} which corresponds to $\sim 1 \times 10^{-7}\ L_{\rm Edd}$ for a $6.5 \times 10^9$~$M_{\odot}$ black hole \citep{ehtm87}. The average radio flux density from V404~Cyg as measured with VLA over the duration of our observation was $\sim 0.5$~mJy, including some flaring activity where the flux density rises to $>1$~mJy. This is consistent with historical averages and the previously noted factor of a few variability seen in long-term radio monitoring \citep{plotkin19}.

\begin{figure*}[ht!]
\epsscale{1.15}
\plotone{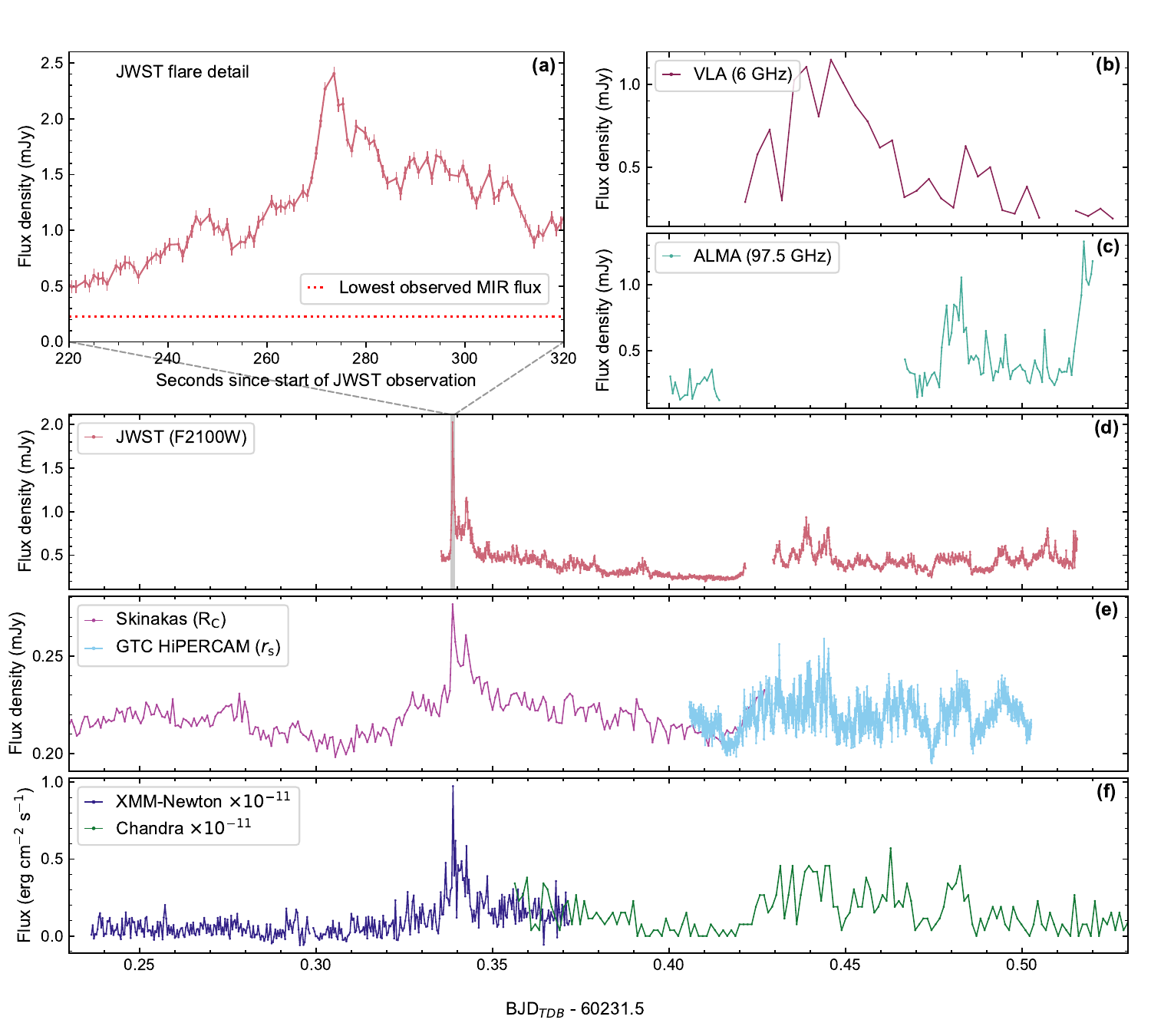}
\caption{Light curves from our multi-wavelength observational campaign. (a) JWST group-level light curve showing detailed view of the large MIR flare. (b) VLA light curve. (c) ALMA light curve. (d) JWST integration-level light curve. (e) Skinakas Observatory and GTC HiPERCAM light curves. (f) XMM-Newton and Chandra light curves. Correlated variability is present throughout, particularly during the large MIR flare and about halfway through the second JWST exposure. 
\label{fig:mwlc}}
\end{figure*}

Intense variability strongly correlated across the electromagnetic spectrum is clearly present in this new dataset. Figure \ref{fig:mwlc} shows the overall multi-wavelength light curves of V404~Cyg. The source was significantly variable in all bands throughout the duration of our observations. There is a large flare apparent $\sim 220$ seconds after the start of the first JWST exposure. This dramatic flare is both large in amplitude and rapid in onset, so it facilitates powerful constraints on the origin of the MIR variability. Notable variability was also present during the second JWST exposure on similar timescales but with lower amplitude.

\begin{figure}[ht!]
\epsscale{1.15}
\plotone{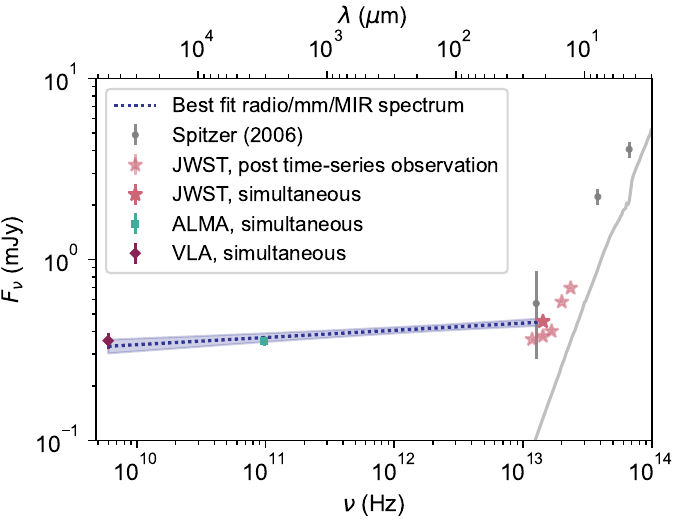}
\caption{The radio to MIR average SED of V404~Cyg including the Spitzer observations from 2006 for comparison. The MIR flux densities have been corrected for interstellar extinction as described in the text. The solid gray line is a model stellar atmosphere representing the secondary star \citep[adopted from][]{hynes09}. The error bars represent statistical uncertainties only, and in most cases are smaller than the data points. The dotted line and shaded region indicate the best fit simultaneous radio/mm/MIR spectral index and 90\% confidence region, respectively. An additional 5\% calibration uncertainty was added in quadrature with the statistical uncertainties to generate the confidence region (based on \citet{2017ApJS..230....7P} for VLA, \citet{cortes_alma_2025} for ALMA, and estimated based on aperture correction uncertainties for JWST) The approximately flat spectrum is consistent with that expected from a jet. 
\label{fig:sed}}
\end{figure}

Figure \ref{fig:sed} shows the spectral energy distribution (SED) of V404~Cyg constructed from our radio, mm, and MIR measurements. We will present and study the full SED in a later work. We correct the observed MIR flux densities for interstellar extinction adopting $A_V = 4.0$ \citep{hynes09, cas93,2021ApJ...916...33G} and $R_V = 3.1$. Considering only the simultaneous portions of the VLA, ALMA, and JWST observations, the radio/mm/MIR spectral slope is $\alpha = 0.04 \pm 0.01$, consistent with the flat or slightly inverted spectrum expected from the optically thick region of a jet. JWST confirms the previously observed excess of MIR emission above that expected from the companion star. The non-simultaneous JWST images taken at the end of the observations show the excess to increase with wavelength. The MIR flux measurements in the three longest wavelength filters (F2550W, F2100W, and F1800W) are still consistent with an approximately flat radio/mm spectrum. The flux measurements in the two shorter wavelength filters (F1500W and F1280W) appear to contain a significant contribution from the donor star, which begins to dominate the SED in this region. However, due to the non-simultaneity of the JWST MIRI images taken after the time-series observations and the significant variability on short timescales, there are some limitations on how much we can infer about the shape of the SED in the MIR.

\section{Discussion}

It is immediately evident by visual inspection of Figure \ref{fig:mwlc} (a) that the rise time of the large flare is short; the flux density increases by about 50\% ($\sim 1.0$~mJy) over just 3.6~seconds. This constrains the size of the emitting region to $\le$3.6~light-seconds, more than an order of magnitude smaller than the $\sim 80$~light-second size of the accretion disk which we adopt from previous studies \citep{hynes09, 2018A&A...620A.110A}.

To examine the possibility of this radiation being thermal in origin, we can calculate the brightness temperature of a blackbody limited to diameter $\le 3.6$~light-seconds. To facilitate a direct comparison, we bin the JWST data to 60~seconds, to match the time resolution of the Skinakas observatory data. A hypothetical blackbody of diameter 3.6~light-seconds producing the dereddened \citep{2021ApJ...916...33G} 21~$\mu$m flux density of $\sim 0.84$~mJy would have a temperature of about $8 \times 10^5$~K. A blackbody of this size and temperature should produce optical (R\textsubscript{C}) flux density of about 1~Jy. The increase in R\textsubscript{C} flux density observed with Skinakas Observatory, after dereddening \citep{1989ApJ...345..245C}, is only approximately 0.9~mJy. This is inconsistent with the hypothetical blackbody by three orders of magnitude, ruling out such an origin for the large flare and leaving synchrotron from a jet as the most likely emission mechanism. 

Throughout the duration of our multi-wavelength observations there were apparent correlations in the emission across all wavelength bands. Detailed timing analyses will be explored in a future study. The first large, double-peaked flare is visible in the data from the three active observatories at that time: XMM-Newton, Skinakas, and JWST. The variability throughout the second JWST exposure is well-correlated with the HiPERCAM and Chandra X-ray observatory data. Most importantly, there is a pronounced feature common between the X-ray, optical, 21~$\mu$m, and millimeter light curves (and perhaps radio with a $\sim$10~minute lag) about halfway through the second JWST exposure (${\rm BJD}_{TDB} - 60231.5 \approx 0.475$). Radio emission from V404~Cyg in quiescence has been well-established to arise from a jet \citep{radiospec}. As the ALMA 97.5~GHz flux density is similar to that measured with the VLA at 6~GHz implying an approximately flat spectrum, it can be expected to also originate in the jet \citep{bkjet}. The correlated variability at similar flux levels observed with JWST and ALMA then suggests that the 21~$\mu$m radiation too is likely to arise from the jet. 

We can make an estimate of the energetics of such a jet by approximating the first dramatic flare as a discrete ejection of synchrotron emitting plasma. One theoretical model which reproduces well the observed flat radio spectra attributed to relativistic jets invokes internal shocks from shells of gas ejected with variable velocities propagating along the jets \citep{2014MNRAS.443..299M}. The collision of faster and slower ejecta leads to shocks which re-energize the adiabatically cooling expanding conical jets, producing the flat spectra. Should we consider the flare to be a discrete ejection in an otherwise steady jet, we can use a simplified model \citep{1994hea..book.....L} to estimate the minimum energy needed to account for the observed monochromatic luminosity. The equations used for these calculations are detailed in Appendix A. To find the volume of the emitting region, we can use the rise time of the flare and assume a modest jet opening angle. Black hole X-ray binary jets are often found to be highly collimated. Two systems with measured jet opening angles in the hard state are the BHXRT MAXI~J1820+070 (opening half-angle $\phi=0.45^{+0.13}_{-0.11}$~deg \citep{2021MNRAS.504.3862T}), and Cygnus~X-1, a persistent black hole X-ray binary ($\phi \sim 0.4 - 1.8$~deg \citep{2019MNRAS.484.2987T}). If we adopt a similar opening angle of 0.8~deg for our calculations and use the observed flux density and radio parallax measured distance to V404~Cyg, we find the minimum energy in the jet to be $\sim 1 \times 10^{34}$~erg. We can also calculate the magnetic field associated with the minimum energy (assuming equipartition between particle and magnetic field energy densities) to be $B_{min}\approx 650$~G. From this magnetic field strength, we can calculate the Lorentz factors of the emitting electrons to be $\gamma_e \approx 140$. These values appear plausible but imply that the timescale for these electrons to cool via synchrotron radiation to be $t_{cool} \approx 13$~seconds, longer than the rise time of the flare. A synchrotron cooling time consistent with the rise time (i.e. electrons lose energy quickly enough to radiate in the MIR) could be realized if the jet opening angle is smaller, some of the energy is in ions rather than electrons, the source components are out of equipartition, or with some combination of these factors. Under the assumption of a standard advection dominated accretion flow model, the X-ray luminosity scales as $L_X \propto \dot{M}^2$ below a state transition luminosity \citep{1995ApJ...452..710N} which occurs at 2\% of Eddington luminosity ($L_{\rm trans}=0.02\ L_{\rm Edd}$) \citep{2003A&A...409..697M}. We can then estimate the maximum available accretion kinetic power to be $\sqrt{L_X L_{\rm trans}}=1.2\times 10^{35}$~erg~s\textsuperscript{-1}. The estimated kinetic power is then less than 10\% of the available kinetic power of the accretion flow, leaving a budget for deviating somewhat from the minimum energy conditions to allow a faster cooling time. The other uncertainties noted are also quite plausible.

We also consider an alternative model \citep{2013MNRAS.430.3196V} which suggests that infrared excesses and fast optical variability correlated with X-rays could originate in an extended hot accretion flow rather than a jet. This model has the advantage of reproducing well the X-ray spectral characteristics of BHXRTs in hard outburst states. In this scenario, similar to the standard model for jets, a combination of synchrotron self-absorption peaks from different zones within a geometrically thick radiatively inefficient accretion flow produces an approximately flat optical to infrared spectrum. This flat spectrum extends down to a turnover frequency (determined by the extent, magnetic field strength, and optical depth of the hot flow) below which the emission drops off sharply. As noted above, one of our key observations is the flat spectrum extending from MIR to radio wavelengths, which is not a feature of this model. As such, if a hot inner flow contributes significantly to the 21~$\mu$m flux density, it would require a coincidence in both frequency and flux for the jet and hot accretion flow components to sum to create the observed flat radio/mm/MIR spectrum. To nonetheless examine the possibility of the observed MIR emission arising from such a flow, we follow the method of \cite{2013MNRAS.430.3196V} using the constraint imposed by the 3.6~second rise time of the large flare. Though the fiducial model applies to hard outburst states, with some assumptions we can extrapolate to quiescent states. If we assume that the luminosity of optically thin flows scales as $L_X \propto \dot{M}^2$, and follow the consideration of this model that the vertical extent and equipartition between magnetic and radiation pressure are independent of accretion rate, the magnetic field strength should scale as $B\propto \sqrt{\dot{M}}$. For this model to apply, the hot flow would need to be an order of magnitude over-magnetized in quiescence. Thus, it is implausible that this model can account for the observed MIR radiation.

Recently, \citet{2025arXiv250523918Z} have presented their investigation of the MIR excess in another LMXB in the Spitzer sample, A0620--00. On the basis of rapid (timescales of $\sim$minutes) variability and a MIR spectral index of $\alpha = 0.72 \pm 0.01$, they also rule out a circumbinary disk as the origin of the MIR excess in that system. While they do not exclude the possibility of a contribution from partially self-absorbed synchrotron radiation, they find that thermal bremsstrahlung from a disk wind can account for the excess. It should be noted that the quiescent X-ray luminosity of A0620--00 ($L_X \approx 10^{-9} L_{Edd}$) is significantly lower than that of V404~Cyg. At the higher luminosity we observe we see a higher ratio of radio/sub-mm flux to MIR making a jet origin more likely. Furthermore, the wind of \citet{2025arXiv250523918Z} was spectroscopically inferred to be warm, at temperatures of a few times $10^4$~K, while we infer a brightness temperature of an order of magnitude higher based on the amplitude of seconds-timescale variability. Such high brightness temperatures are more consistent with a synchrotron origin.

\section{Conclusions}

Here we have presented the results of simultaneous multi-wavelength observations of the quiescent BHXRT V404~Cygni. These are among the first JWST observations of a quiescent black hole and the first time V404~Cyg has been observed with JWST or ALMA. The MIR excess observed in 2005 with Spitzer was hypothesized to arise from 1) a blackbody-like source (i.e. circumbinary material or the accretion disk), or 2) an unresolved relativistic jet. The time-domain nature of our study was essential to resolving this long-standing question. The large amplitude 21~$\mu$m variability on short timescales is inconsistent with a thermal blackbody-like emitter, ruling out circumbinary dust or viscously heated accretion disk origins and leaving synchrotron emission from the inner jet as the most likely production mechanism. Correlated variability across all observed wavelengths, especially the mm band, also supports the jet origin of the MIR excess at and longward of 21~$\mu$m. Additionally, the measured 21~$\mu$m flux density is consistent with an approximately flat spectrum as extrapolated from the simultaneous radio and mm observations, as is expected from partially self-absorbed synchrotron radiation from a jet. Thus, this is the first compelling detection of an MIR jet in a quiescent BHXRT. However, it should be noted that we cannot rule out a small thermal contribution to the MIR excess at wavelengths short of 21~$\mu$m. Our results reassert the persistence of relativistic jets into quiescence and demonstrate similarity between the quiescent and hard outburst states of black hole X-ray binaries. These results show that JWST can be used to study low-luminosity jets in quiescent BHXRTs, probing the regions closest to the black hole. Constraining jet parameters in BHXRTs with JWST can furthermore inform the models used to explain observations of nearby quiescent supermassive black holes. 

\begin{acknowledgments}

E.S.B. acknowledges support from the Louisiana Space Grant Consortium (LaSPACE) Graduate Student Research Assistance (GSRA) program. 
A.J.T. and Q.H. acknowledge that this research was undertaken thanks to funding from the Canada Research Chairs Program, the Natural Sciences and Engineering Research Council of Canada (NSERC; funding reference number RGPIN-2024-04458), and the Canadian Space Agency (CSA; funding reference number 23JWGO2B08). R.M.P. acknowledges support from NASA under award No. 80NSSC23M0104. T.D.R. is an INAF IASF research fellow. T.S. acknowledges support from the Agencia Estatal de Investigación del Ministerio de Ciencia, Innovación y Universidades (AEI-MCIU) under grants PID2020-114822GB-I00 and PID2023-151588NB-I00. C.O.H. is supported by NSERC Discovery Grant RGPIN-2023-04264. G.R.S. is supported by NSERC Discovery Grant RGPIN-2021-0400. 

This work is based in part on observations made with the NASA/ESA/CSA JWST. The data were obtained from the Mikulski Archive for Space Telescopes at the Space Telescope Science Institute, which is operated by the Association of Universities for Research in Astronomy, Inc., under NASA contract NAS 5-03127 for JWST. The specific observations analyzed can be accessed via \dataset[DOI: 10.17909/7p5n-8888]{https://doi.org/10.17909/7p5n-8888}. These observations are associated with program \#3384. Support for E.S.B. and R.I.H. under program \#3384 was provided by NASA through grant JWST-GO-03384.001-A from the Space Telescope Science Institute, which is operated by the Association of Universities for Research in Astronomy, Inc., under NASA contract NAS 5-03127. 

This paper makes use of the following ALMA data: ADS/JAO.ALMA\#2023.1.01719.S. ALMA is a partnership of ESO (representing its member states), NSF (USA) and NINS (Japan), together with NRC (Canada), NSTC and ASIAA (Taiwan), and KASI (Republic of Korea), in cooperation with the Republic of Chile. The Joint ALMA Observatory is operated by ESO, AUI/NRAO and NAOJ. The National Radio Astronomy Observatory is a facility of the National Science Foundation operated under cooperative agreement by Associated Universities, Inc. 

This work is based in part on observations made with the Gran Telescopio Canarias (GTC), installed at the Spanish Observatorio del Roque de los Muchachos of the Instituto de Astrofísica de Canarias, on the island of La Palma under Director's Discretionary Time (program GTC07-23BDDT). This work is based in part on data obtained with the instrument HiPERCAM, built by the Universities of Sheffield, Warwick and Durham, the UK Astronomy Technology Centre, and the Instituto de Astrofísica de Canarias. Development of HiPERCAM was funded by the European Research Council, and its operations and enhancements by the Science and Technology Facilities Council. 

This work is based in part on observations made by the Chandra X-ray Observatory under Director's Discretionary Time (Proposal ID 24408930), contained in the Chandra Data Collection ~\dataset[DOI: 10.25574/cdc.446]{https://doi.org/10.25574/cdc.446}. We are grateful to the Director and observatory staff for facilitating these observations.

The scientific results reported in this article are also based in part on observations obtained with XMM-Newton as a Target of Opportunity program (Revolution 4367 ObsID 0932390301). XMM-Newton is an ESA science mission with instruments and contributions directly funded by ESA Member States and NASA. We are grateful to the Project Scientist and the SOC for facilitating these observations. 

This research has made use of the Astrophysics Data System, funded by NASA under Cooperative Agreement 80NSSC21M00561.

\end{acknowledgments}

\appendix

\section{Jet energetics}
\subsection{Minimum energy requirements}
Considering the large MIR flare as a discrete ejection of synchrotron emitting plasma, we follow the method outlined in \citet{1994hea..book.....L} and apply a simplified approach to estimating the minimum energy requirements to produce the observed synchrotron radiation in a jet. The assumptions of this model include 1) source components are close to equipartition, 2) $\eta=1$ (i.e., all energy in electrons, none in protons or nuclei), 3) particles and magnetic field fill source volume uniformly, and 4) the emission region is optically thin (spectral index $\alpha=-0.75$, where flux density per frequency $F_\nu \propto \nu^\alpha$). Should the MIR emission arise in a region which is partially optically thick, this method instead gives us a lower limit on the minimum energy. Under these assumptions (\citealt[][eq. 19.29]{1994hea..book.....L}),
\begin{displaymath}
    E_{min} \approx 3.0\times10^6\ \eta^{4/7}\ V^{3/7}\ \nu^{2/7}\ L_{\nu}^{4/7}
\end{displaymath}
in SI units, where $\eta=(1+\beta)$ and $\beta$ is the fraction of energy stored in protons rather than electrons, $V$ is the volume of the emitting region, $L_\nu$ is the monochromatic luminosity, and $\nu$ is the frequency.
\subsection{Magnetic field corresponding to minimum energy}
The strength of the magnetic field associated with the minimum energy requirements under the equipartition assumption can be calculated as (\citealt[][eq. 19.30]{1994hea..book.....L}):
\begin{displaymath}
    B_{min} = 1.8 \left( \frac{\eta L_{\nu}}{V} \right)^{2/7} \nu^{1/7}
\end{displaymath}

The Lorentz factor of electrons emitting synchrotron radiation at that frequency can be estimated as (generalized from \citealt[][eq. 9.5]{xrbjets}):
\begin{displaymath}
    \gamma_e \approx 0.3 \left( \frac{\nu}{10^9} \right)^{1/2} B^{-1/2}
\end{displaymath}

The time for these electrons to cool via synchrotron radiation can be calculated by dividing their internal energy by the power radiated:
\begin{displaymath}
    t_{cool}=\frac{\gamma_e m_e c^2}{2 \sigma_T U_{mag} \gamma_e^2 \sin^2(\theta)}
\end{displaymath}
in SI units, where $m_e$ is the electron mass, $\sigma_T$ is the Thomson cross section for an electron, $U_{mag}$ is the magnetic field energy density, and $\theta$ is the pitch angle. After averaging over an isotropic distribution of pitch angles, and substituting $U_{mag}=B^2 / 2\mu_0$, this can be simplified to:
\begin{displaymath}
    t_{cool}= \left( m_e c^3 \right) \left( \frac{2}{3} \sigma_T \frac{B^2}{\mu_0} v^2 \gamma_e \right)^{-1}
\end{displaymath}
where $\mu_0$ is the vacuum permeability and $v$ is velocity.

\bibliography{v404cyg_qjet_b}{}
\bibliographystyle{aasjournalv7}

\end{document}